# AUTOMATIC FREEWAY BOTTLENECK IDENTIFICATION AND VISUALIZATION USING IMAGE PROCESSING TECHNIQUES

**Hao Chen**
Virginia Tech Transportation Institute
3500 Transportation Research Plaza, Blacksburg, VA 24061
Phone: (540) 231-0254 Fax: (540) 231-1555
hchen@vtti.vt.edu

**Hesham A. Rakha (Corresponding author)**
Charles E. Via, Jr. Department of Civil and Environmental Engineering
Virginia Polytechnic Institute and State University
3500 Transportation Research Plaza, Blacksburg, VA 24061
Phone: (540) 231-1505 Fax: (540) 231-1555
hrakha@vt.edu

**ABSTRACT**
This paper develops an automatic freeway bottleneck identification and visualization algorithm using a combination of image processing techniques and traffic flow theory. Unlike previous studies that are based solely on loop detector data, the proposed method can use traffic measurements from various sensing technologies. Four steps are included in the proposed algorithm. First, the raw spatiotemporal speed data are transformed into binary matrices using image binarization techniques. Second, two post-processer filters are developed to clean the binary matrices by filtering scattered noise cells and localized congested regions. Subsequently, the roadway geometry information is used to remove the impact of acceleration zones downstream of bottlenecks and thus locate bottlenecks more precisely. Finally, the major characteristics of bottlenecks including activation and deactivation points, shockwave speeds and traffic delay caused by bottleneck are automatically extracted and visualized. The proposed algorithm is tested using loop detector data from I-5 demonstrating that the proposed method outperforms the state-of-the-art methods for congestion identification. The second test using INRIX data from I-66 demonstrates ability of the proposed algorithm to accurately extract and visualize bottleneck characteristics.

**INTRODUCTION**

Congestion has proven to be a serious problem across urban areas in the United States. Traffic congestion reduces the utilization of the transportation infrastructure and increases traveler travel times, air pollution, and fuel consumption levels. In 2007, congestion cost highway users 4.2 billion extra hours of sitting in traffic and an extra 2.8 billion gallons of fuel. This all translated into an additional $87.2 billion in congestion costs for road users in 2007, which represented a 50% increase in congestion costs compared to data from the previous decade. Even though the recent economic downturn is said to have marginally eased the congestion problem nationwide, new evidence shows an uptrend of traffic and, consequently, congestion is back [1]. Tackling congestion (both recurrent and non-recurrent) has proven to be a challenge for highway agencies. Adding capacity in response to congestion is becoming less of an option for these agencies due to a combination of financial, environmental, and social constraints. Consequently, the main focus has been on improving the performance of existing facilities through continuous monitoring and dissemination of traffic information [2].

Freeway bottlenecks are important contributors to congestion. Unlike traffic congestion, freeway bottlenecks are usually caused by specific physical conditions, such as lane drop, merge, diverge or weaving sections, etc. They can also be caused by temporary situations, such as incidents [3]. The congestion formed upstream of traffic bottlenecks is a complex spatiotemporal nonlinear dynamic process. One goal of Advanced Traffic Management Systems (ATMSs) within the Intelligent Transportation Systems (ITSs) is to identify bottlenecks within the transportation system and then take actions to alleviate bottlenecks. Correctly identifying freeway bottlenecks is critical in understanding traffic dynamics under congested conditions and characterizing the spatiotemporal interactions and correlations that exist along roadway segments. By diagnosing the traffic dynamics and spatiotemporal interactions caused by freeway bottlenecks, transportation agencies can develop appropriate solutions to mitigate congestion and improve the performance of a freeway network [4]. In summary, the results of bottleneck identification are beneficial in predicting the propagation of bottlenecks into the future, understanding bottleneck causes, and identifying bottleneck mitigation strategies.

Generally, freeway bottlenecks are difficult to identify given that bottlenecks are very complex traffic phenomena. More importantly, the spatiotemporal evolution of congestion upstream of bottlenecks is extremely difficult to quantify, since the spatiotemporal evolution varies from one day to another [5]. For example, the spatiotemporal location of active traffic bottlenecks on workdays may differ significantly from active traffic bottlenecks on weekends. In addition, bottlenecks during the morning peak are not the same as traffic bottlenecks during the evening peak.

During the past several decades, researchers have developed various methods to identify freeway bottlenecks. Chen et al. [6] developed an automatic bottleneck identification algorithm to extract bottleneck locations, activation times, and the corresponding delays caused by each bottleneck. Three parameters are used in Chen et al.'s algorithm, which are the data aggregation interval, maximum upstream speed, and minimum speed differential between neighboring detectors. Wieczorek et al. [7] extended the work in [6] to identify the optimum parameters and display the active bottleneck features using graphical tools. Using loop detector data along I-5 in the Portland Area, the optimum settings of 3 minute-aggregation, 40 mph maximum upstream speed and 15 mph speed differential were obtained for the selected freeway corridor. A similar algorithm to [6] was proposed by Bai et al. [8] to automatically identify bottleneck locations, activation and deactivation times using loop detector data. The criteria in the algorithm were developed using three parameters, including data aggregation level, maximum occupancy measured by the upstream detector and maximum occupancy differential between neighboring detector pair. Ban et al. [9] proposed a method for bottleneck identification and calibrated the model using simulation. The identification was conducted using percentile speeds based on data from multiple days to extract the spatiotemporal bottleneck influence area. The experimental results demonstrated the accuracy of the proposed method in bottleneck identification and the improvement of calibration procedure for the current practice of simulation.

In the previous several decades, loop detectors were widely used to collect traffic data. However, the maintenance cost for loop detectors are generally very high and repairing the malfunction detectors usually result in damages on road surface [2]. In the recent years, many transportation agencies start to replace loop detector data by probe data., Travel times or average speeds are usually collected for each roadway segment by probe vehicles. Private company such as INRIX has national wide segment-based speed data all over United States, and the collected probe data is supplemented by traditional road sensors and other sources [10]. Moreover, traffic data collected by more advanced sensing technologies, such as mobile or Bluetooth devices, and participatory sensing data from social media, are rapidly growing and potentially could be very good data sources for analyzing bottleneck. Unlike most of the previous studies using loop detector data as model input, the bottleneck identification method developed in this paper uses spatial temporal speed data as input and it's flexible to use traffic measurements from various traffic sensing technologies such as loop detector, INRIX probe, mobile or Bluetooth devices or participatory sensing data, given that these traffic sensing data can be used to generate spatial temporal speed data. On the other hand, previous studies usually extract partial bottleneck characteristic information. In this paper, a more comprehensive bottleneck characteristics including activation and deactivation points, shockwave speed, and traffic delay cause by each bottleneck can be automated extracted and visualized using the proposed method. The extracted bottleneck characteristics and the visualization of bottleneck can potentially very helpful for transportation agencies to analyze features of bottleneck, find the causes for bottleneck and eventually develop effective strategies to alleviate bottleneck.

This paper develops an automatic freeway bottleneck identification algorithm using image processing techniques. In the proposed framework, the raw spatiotemporal speed data are transformed into binary matrices by image binarization technique. Then two post filers are proposed to further clean up binary results by filtering noise cells and localized congested regions. Subsequently, roadway geometry information is used to remove the impact of acceleration area downstream of bottleneck. On the last step, the major characteristics of bottleneck including activation and deactivation points, shockwave speed and traffic delay caused by bottleneck are extracted and visualized. Two case studies using loop detector data on I-5 and INRIX probe data on I-66 are conducted. The test results demonstrate the proposed method can accurately and effectively identify bottlenecks.

The remainder of this paper is organized as follows. Firstly, the proposed methodology framework for bottleneck identification is introduced. Followed by the description of all the image processing techniques used in the proposed method. Subsequently, the results of case study are presented to demonstrate the proposed method can effectively and accurately identify bottlenecks. Finally, conclusions and recommendations for future work are provided.

**METHODOLOGY**
**The Framework of Bottleneck Identification**

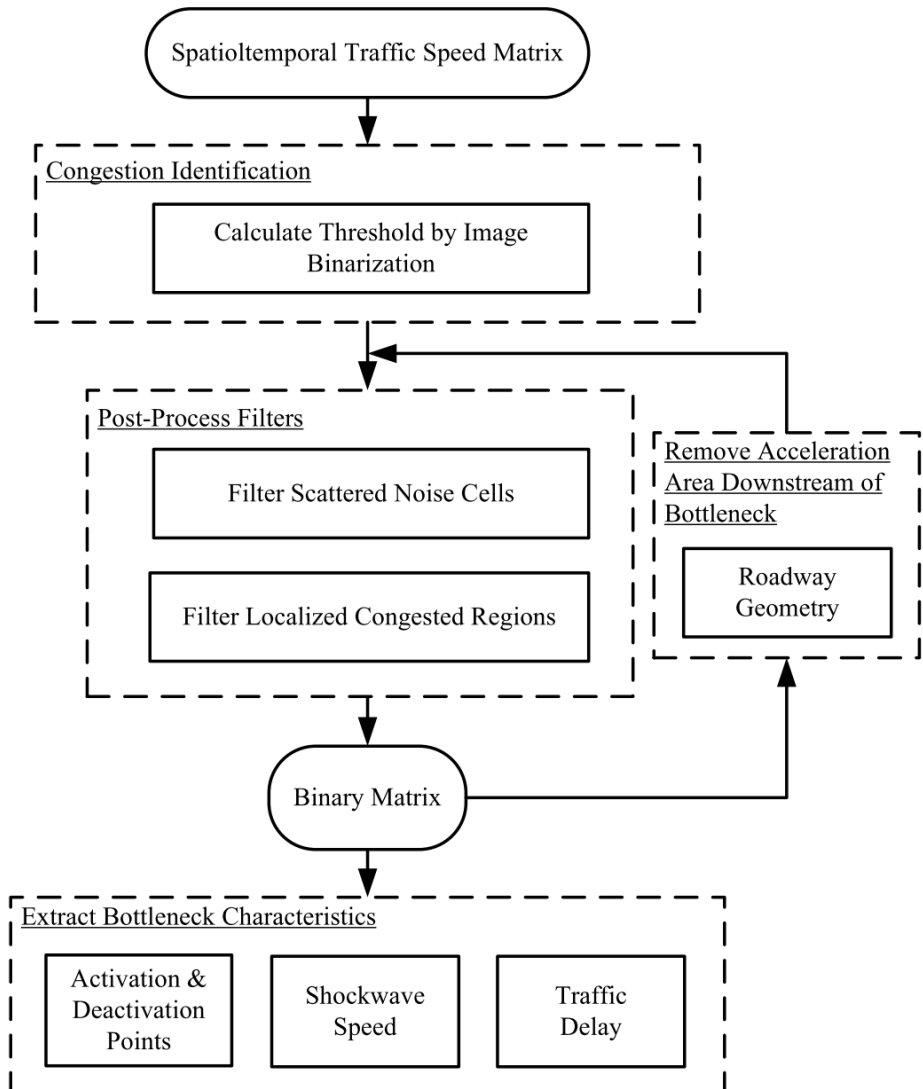

**FIGURE 1. The proposed framework of bottleneck identification.**

FIGURE 1 presents the proposed framework of bottleneck identification. The input data in the framework is the spatial and temporal speed matrix generated from INRIX probe data. While the proposal framework uses speed data from INRIX, it's not constrained to other data sources such as loop detector, Bluetooth or mobile sensing data, given that these sensing data can also be used to generate spatial temporal speed matrix.

    Initially, an image binarization method is implemented on speed data to find the optimum threshold to differentiate congested and uncongested traffic conditions. In this way, each daily speed matrix is transformed into a binary matrix with only 0 and 1 values. Thereafter, two post-process filters are proposed to clean up the binary matrix. Specifically, mathematical morphology operations are used to eliminate scattered noise cells, and a connected area analysis is implemented to filter localized congested regions. After applying the post-process filters, the impact of acceleration area downstream of bottleneck is removed by using roadway geometry information. The post-process filters are implemented again on the modified speed matrix after removing the impact of acceleration area. The purpose of this step is to further separate neighboring bottlenecks which may be connected together as a single bottleneck. Subsequently,

each individual bottleneck area can be identified and the corresponding bottleneck characteristics including activation and deactivation points, shockwave speed and traffic delay caused by bottleneck are extracted. The details of each step of the proposed algorithm are presented in the following sections.

**Congestion Identification by Image Binarization**
The purpose of congestion identification is to find a speed threshold to differentiate speed value into binary result, either 1 (congested traffic condition) or 0 (uncongested traffic condition). Many studies discover that traffic speed data follow a bimodal distribution, such as lognormal mixture distribution represented in (1) [4, 11, 12]. An illustration of using two distributions to fit field speed data is presented in FIGURE 2, the field data were collected by INRIX for an I-66 freeway corridor on June 19, 2012.

$$f(u|\lambda,\mu_1,\mu_2,\sigma_1,\sigma_2) = \lambda \frac{1}{\sqrt{2\pi}u\sigma_1} e^{\frac{(\ln u - \mu_1)^2}{2\sigma_1^2}} + (1-\lambda)\frac{1}{\sqrt{2\pi}u\sigma_2} e^{\frac{(\ln u - \mu_2)^2}{2\sigma_2^2}} \tag{1}$$

where $(\mu_1,\sigma_1)$ and $(\mu_2,\sigma_2)$ are the mean and standard deviation of the first and second component distributions and $\lambda$ is the mixture parameter.

The process to separate congested and uncongested traffic data is very similar to image binarization, in which the gray level image is transformed into a binary image with 1 or 0 for each pixel. The INRIX speed measurements along a freeway corridor for a certain time period can be processed into a spatial temporal speed matrix. So the speed data collected by an entire day can be represented by a speed matrix or an image. If the daily speed matrix is considered as a gray level image, then image binarization can be used for congestion identification. The Otsu's method is a well-established image binarization technique and is selected in this section by considering its simplicity and fast computation [13]. More importantly, the Otsu's method finds a global threshold to separate two data classes in an image and it produces relatively good performance if the histogram can be assumed to have bimodal distribution with a deep valley between two peaks [14]. Therefore, the Otsu's method is very appropriate to solve the traffic congestion identification problem. Note that other similar or more advanced image binarization techniques such as the Niblack method [15], the Bernsen method [16] may also be utilized for congestion identification.

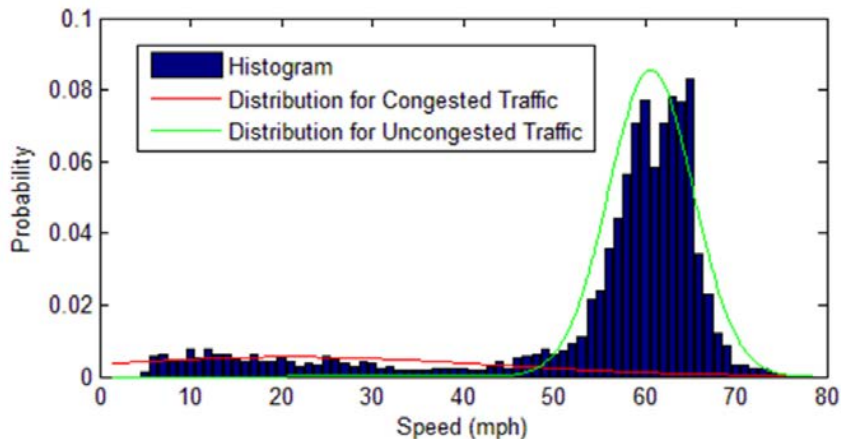

**FIGURE 2. Two distributions for traffic speed field data.**

The Otsu's method aims to find the global threshold that minimizes the weighted within-class variance of the data matrix. The algorithm assumes that the image can be divided into two classes of pixels and then calculates the optimum threshold for classification so that the combined spread is minimal. This process of minimizing the within-class variance also entails maximizing the between-class variance. Considering our problem of identifying congestion using spatiotemporal traffic speed data, this approach can maximize the variance between two classes (i.e. between congested and uncongested traffic conditions). The main advantages of this method are that no pre-assigning of parameters is required and the calculation is conducted directly on the gray level histogram, so the computation speed is very fast. The equations in the Otsu's method are presented as below.

Assume the threshold value is $t$, the weighted within-class variance $\sigma_w(t)$ is computed as

$$\sigma_w^2(t) = q_1(t)\sigma_1^2(t) + q_2(t)\sigma_2^2(t) \tag{2}$$

where $q_1(t)$ and $q_2(t)$ are the probabilities of the two classes separated by the threshold $t$, and $\sigma_1(t)$ and $\sigma_2(t)$ are the variances of two classes. The class probabilities are computed by the histograms as below.

$$q_1(t) = \sum_{i=1}^{t} P(i), \quad q_2(t) = \sum_{i=t+1}^{I} P(i) \tag{3}$$

The class means $\mu_1(t)$ and $\mu_2(t)$ are given by

$$\mu_1(t) = \sum_{i=1}^{t} \frac{iP(i)}{q_1(t)}, \quad \mu_2(t) = \sum_{i=t+1}^{I} \frac{iP(i)}{q_2(t)} \tag{4}$$

The individual class variances are computed as

$$\sigma_1^2(t) = \sum_{i=1}^{t} [i - \mu_1(t)]^2 \frac{P(i)}{q_1(t)}, \quad \sigma_2^2(t) = \sum_{i=t+1}^{I} [i - \mu_2(t)]^2 \frac{P(i)}{q_2(t)} \tag{5}$$

Here, the speed data are normalized between 0 and 1. Thus, the optimum threshold $u^t$ corresponds to the value of $t$ that minimizes (2).

The speed threshold is usually close to the value of speed at capacity, given that speed at capacity is the cut-off point from free flow to congested traffic. However, if the speed matrix only includes one class of traffic condition (either uncongested or congested traffic condition) instead of two classes of conditions, then the speed threshold by the Otsus's method may be computed as a very small or large value. Two parameters $\theta_1$ and $\theta_2$ are introduced here to identify this situation. Considering the speed at capacity approximately ranges from 0.7 to 0.8 of free flow speed for freeway, the values of two parameters are set as $\theta_1=0.3$ and $\theta_2=0.85$ and the free flow speed value $u_{free}$ can simply set as the speed limit of freeway corridor. If the calculated speed threshold is either too small or too large, this means the data only fall into one class of either congested or uncongested condition and so the threshold from previous congested days (denoted by $u_{pre}$) or a predefined value such as 0.75 of speed limit can be used. The above mentioned process is represented in (6).

$$u^* = \begin{cases} u^t & \theta_1 \cdot u_{free} \leq u^t \leq \theta_2 \cdot u_{free} \\ u^{pre} & otherwise \end{cases} \tag{6}$$

FIGURE 3 demonstrates the result of congestion identification using image binarization technique. The speed contour is presented in FIGURE 3 (a), and the binary result by using the Otsu's algorithm is presented in FIGURE 3 (b). In the binary matrix, the red and blue colors are used instead of black and white colors. The blue color denotes uncongested traffic condition and the red color represents congested condition. Thereafter, two post-process filters are proposed to refine the binary result. The adjusted binary results after applying two filters to clear up the image are presented in FIGURE 3 (c) and (d)

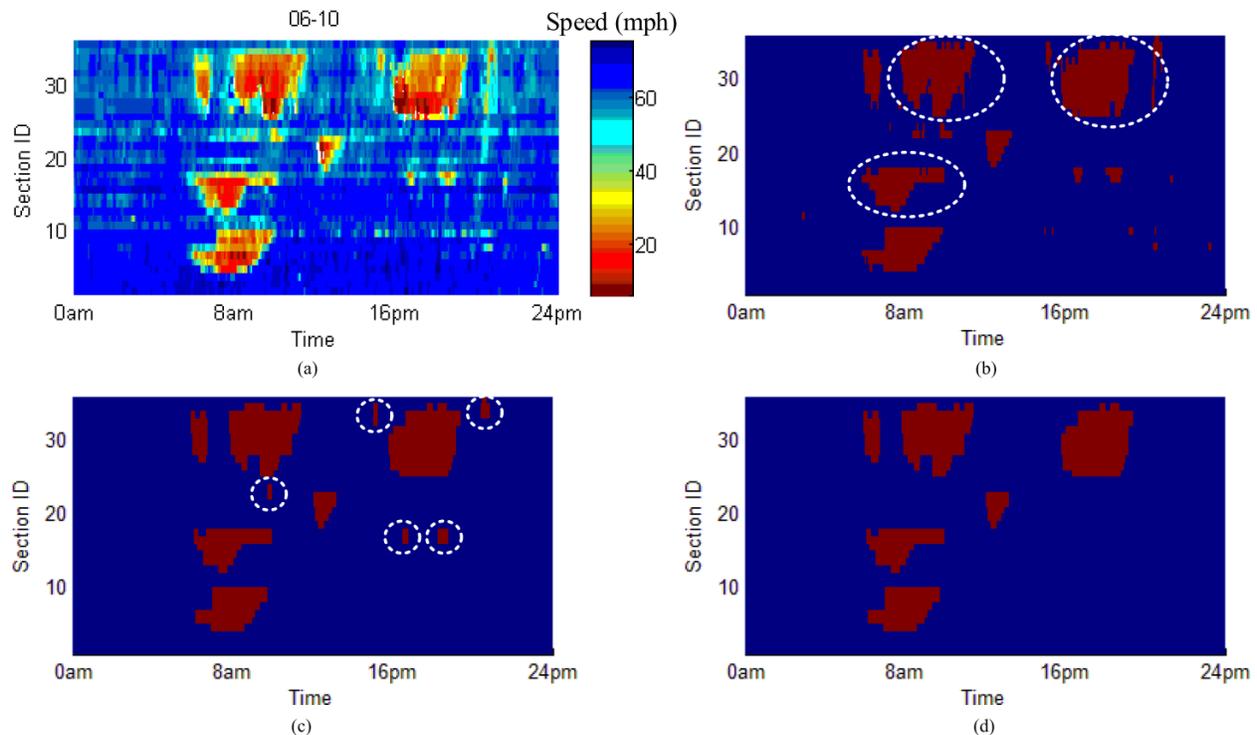

**FIGURE 3.** Congestion identification. (a) speed contour. (b) binary result using the Otsu's method. (c) adjusted binary image after filtering scattered noise cells. (d) adjusted binary image after filtering localized congested regions.

**Post-Process Filters**

Binary matrix with 0 and 1 values can be obtained from congestion identification. However, several problems are observed in the binary image of FIGURE 3 (b), considering the purpose of bottleneck identification. Frist, many scattered cells or pixels are recognized as congested traffic condition, which include some isolated small areas and some pixels connected to the edges for large congested regions. But it may not be appropriate to recognize these scatter cells as traffic bottleneck, given that their sizes are very small and they don't have major impact over spatial and temporal as a bottleneck. Second, some scattered cells inside of bottlenecks are incorrectly recognized as uncongested traffic condition, which doesn't meet the physical law of traffic. The cause for this issue can be either 1) the binarization method doesn't work very well for this region, or 2) the raw speed data have errors. The latter case is actually very common since the INRIX data sometimes have missing data problem and historical information are used to fill the missing data, which may cause the data inconsistency issue in the bottleneck area. Third, some localized congested regions have very limited impacts for the freeway corridor, given that the sizes of congested regions are negligible by comparing to the major bottlenecks in the corridor.

    The first two issues in the congestion identification binary results are very similar to the salt-and-pepper noise problem in image binarization. The salt noise pixels in a binary image are the small undesirable foreground pixels, and the pepper noise pixels are the small background color holes in the binary image [17]. Mathematical morphology (MM) is an effective approach to remove salt-and-pepper noise [18], thus morphology operations are used to eliminate scattered noise cells in the binary image. Afterward, a connected components analysis is implemented to further remove localized congested regions.

*Filter Scattered Noise Cells*

Mathematical morphology is a theory to analyze and solve the problems related to geometrical structures [18]. The basic idea in binary morphology is to provide an image with a simple, pre-defined shape, drawing conclusions on how this shape fits or misses the shapes in the image. The simple probe is called the structuring element. Assume $A$ is a binary image and $B$ is a structuring element, the basic morphology operators include erosion and dilation, as shown in (7) and (8) respectively.

$$A \otimes B = \bigcup_{b \in B} A_{-b} \tag{7}$$

$$A \oplus B = \bigcup_{b \in B} A_{b} \tag{8}$$

The opening of $A$ by $B$ is obtained by the erosion of $A$ by $B$, followed by dilation of the result by $B$. The closing of $A$ by $B$ is obtained by the dilation of $A$ by $B$, followed by erosion of the result by $B$. The equations are presented as

$$A \circ B = (A \otimes B) \oplus B \tag{9}$$

$$A \bullet B = (A \oplus B) \otimes B \tag{10}$$

In this filter, the morphology opening is firstly used to remove the scattered noise cells that are incorrectly recognized as congestion condition. These noises cells are the salt noise pixels in the binary image. Afterward, the morphology closing is used to remove the scattered noise cells inside of bottlenecks that are incorrectly recognized as uncongested condition. These noise cells can be categorized as the pepper noise pixels in the binary image. FIGURE 3 (b) illustrate the scattered noise cells, in which the salt noise pixels are the isolated "congested" small areas and some pixels connected to the edges for large congested regions (highlighted by white dotted circles), and the pepper noise pixels are the small holes inside of large congested regions. A structure of α1 sections by α2 time intervals is used in the morphology opening and closing operations. Considering the resolution of traffic speed data in, the parameters are set as α1 = 2 and α2 = 3. The modified binary image by morphology operations is presented in FIGURE 3 (c), we can see that most of the scattered noise cells are successfully removed.

*Filter Localized Congested Regions*

After filtering scattered noise cells in the binary image, some localized congested regions highlighted in white dotted circles still exist in the binary matrix as shown in FIGURE 3 (c). The issue can be solved by the connected components analysis [19]. Connected components with fewer than the predefined $α_3$ number of cells are removed from the binary image. The value of parameter $α_3$ can be customized by transportation agencies based on their criterion to define localized congested region. The parameter value is set as $α_3 = 20$, and the adjusted binary image is presented in FIGURE 3 (d). The result indicates the localized congested regions in the dotted circles are effectively removed.

**Remove Acceleration Area Downstream of Bottleneck**

By using the proposed congestion identification method, the acceleration area downstream of bottleneck is usually incorrectly identified as congested region, since traffic speeds in the acceleration area are usually lower than the threshold. This is one of the disadvantages of using speed data to identify congestion instead of traffic flow data. When vehicles pass the bottleneck, it will take some time for vehicles to accelerate given the constraint of vehicle dynamics. This is the reason that the area downstream of bottleneck still shows lower traffic speed. But the acceleration area downstream of bottleneck should not be identified as a part of the bottleneck since vehicles are free to accelerate in this area. We propose to use roadway geometry information to solve this issue. Note that the roadway geometry information include the locations of on/off ramps, lane drop, weaving sections or any other geometry features can result in back-propagation static bottlenecks. By plotting the INRIX data used in this study, we find out the acceleration areas incorrectly recognized as congested regions are usually happened at several specific on-ramp locations. So the following steps are proposed to automatically identify acceleration area downstream of bottleneck.

- Step 1: extract binary matrix for each individual bottleneck (remove other bottleneck regions by assigning values of 0).
- Step 2: extract speed values in the bottleneck region by calculating the dot product of original speed matrix and the binary matrix obtained from step 1.
- Step 3: calculate the vertical projection curve using the speed matrix obtained from step 2.
- Step 4: identify the downstream location of ramp as acceleration area under the following condition

$$\text{ProCurv}(\text{RampLoc}+1) > \lambda_1 \cdot \text{ProCurv}(\text{RampLoc}) \tag{11}$$

Where *ProCurv* represents the vertical projection curve obtained from step 3; *RampLoc* denotes the pre-defined on-ramp locations along the selected I-66 corridor. Here, $\lambda_1$ can be customized by traffic agencies, to apply different levels of sensitivity. A value of $\lambda_1=1.3$ is used in the case study of INRIX data along I-66.

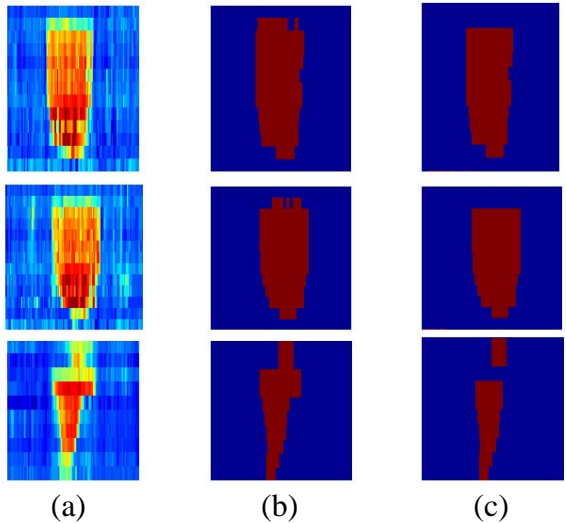

(a)   (b)   (c)

**FIGURE 4. Examples of using roadway geometry information to locate acceleration area downstream of bottleneck. (a) speed contour. (b) binary image before removing acceleration area. (c) binary image after removing acceleration area.**

FIGURE 4 demonstrates examples of using roadway geometry information to locate acceleration area downstream of bottleneck. FIGURE 4 (a) illustrate the speed contour plots of three bottlenecks. The binary results before removing acceleration area are presented in FIGURE 4 (b). And the adjusted binary results after removing acceleration area are presented in FIGURE 4 (c). The results clearly show the improvements of congestion identification results after removing acceleration area. It should be noted that the process of removing acceleration area is also helpful for the problem that multiple bottlenecks are connected together as a single bottleneck. As shown in the methodology framework in FIGURE 1, the post-process filters are implemented again on the modified speed matrix after removing the impact of acceleration area. The purpose of this step is to further separate neighboring bottlenecks, and most of the connected bottlenecks can be effectively separated after this process so that isolated bottlenecks can be obtained.

**Extract Bottleneck Characteristics**
The isolated bottleneck is obtained from the above mentioned procedures. In this section, each bottleneck is processed separately to extract major bottleneck characteristics including bottleneck activation and deactivation points, shockwave speed, and delay caused by bottleneck. The extracted bottleneck characteristics and the visualization of bottleneck can potentially very helpful for transportation agencies to analyze features of bottleneck, find the causes for bottleneck and eventually develop effective strategies to alleviate bottleneck.

*Bottleneck Activation and Deactivation Points*
Activation and deactivation points are located on the front and back of bottlenecks. Here, the front of bottleneck is used as an example for identifying activation and deactivation points. The same process can be used to extract activation and deactivation points on the back of bottlenecks.

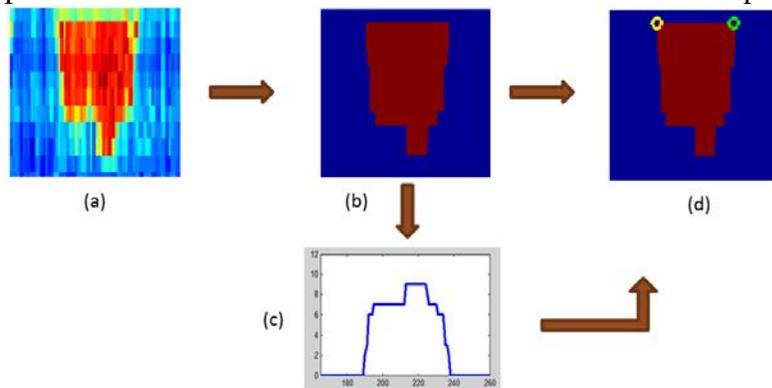

**FIGURE 5. Illustration to locate bottleneck activation and deactivation points; (a) speed contour; (b) binary result from congestion identification; (c) horizontal projection curve; (d) activation and deactivation points.**

Both of time and location information need to be identified for bottleneck activation and deactivation points. A vertical projection curve of the bottleneck's binary matrix can be used to find the front location. Activation and deactivation time points can be located using a horizontal projection curve, as demonstrated in FIGURE 5. The binary result can be used to generate a horizontal projection curve as shown in FIGURE 5 (c). The bottleneck activation time is the first time that the curve value changes from zero to non-zero. Similarly, the bottleneck deactivation time is the last time where the curve value changes from non-zero to zero. The two points are represented in FIGURE 5 (d) by yellow and green colors respectively. Note that the same process can be used to extract activation and deactivation points on the back of bottlenecks. Thus, the activation and deactivation points for each bottleneck can be accurately extracted using the proposed approach.

*Shockwave Speed*

The shockwave speed of each bottleneck can be calculated using bottleneck activation and deactivation points on the front or rear location of bottleneck. The similar method in the previous section can be used to location the rear location of bottleneck by using vertical projection curve of binary matrix. Therefore, the activation points on the front or rear of bottleneck can be used to compute the shockwave speed. It should be noted that shockwave speeds calculated in this manner may deviate from what can be observed in the field. Specifically, shockwave speeds directly calculated from speed matrix can sometimes be unrealistic large, infinity, or negative values. These incorrect speed values may be caused by traffic data measurement errors, data resolution discrepancies, or congestion/bottleneck identification errors. Theoretically, shockwave speeds should be constrained by the fundamental diagram. Here, a more realistic fundamental diagram – Van Aerde traffic steam model [20, 21] is considered to calculate the maximum shockwave speed. The shockwave speed is dynamically changing on the different location in fundamental diagram. The maximum value can be located on the jam density by

$$w_s = -\left[\left(\frac{k_j}{q_c} - \frac{u_f}{u_c^2}\right) + \frac{(u_f - u_c)^2}{u_f u_c^2}\right]^{-1} \tag{12}$$

where $u_c$ is speed at capacity; $u_f$ is free flow speed; $q_c$ is flow at capacity and $k_j$ is jam density. In this manner, shockwave speed needs to be constrained by maximum speed calculated from the fundamental diagram.

*Traffic Delay*
The totally traffic delay for each bottleneck is an important criterion to evaluate bottleneck impacts. Free flow speed can be assumed for areas with no bottleneck. For the cell of roadway section $i$ at time interval $t$, traffic delay can be calculated by

$$d_i(t) = L_i \times q_i(t) \times \left(\frac{1}{u_i(t)} - \frac{1}{u_f}\right) \tag{13}$$

where $L_i$ is segment length; $q_i(t)$ is the traffic volume; $u_i(t)$ represents measured speed. Consequently, the total delay caused by this bottleneck can be calculated by the summation of delay from each cell within the bottleneck as

$$Total\ Delay = \sum d_i(t) \tag{14}$$

Note that there are two ways to obtain traffic volume in (13). The traffic volume can be directed measured by loop detector. If measured traffic volume is not available, the fundamental diagram such as the Van Aerde model in (15) and (16) can be used to estimate traffic volume by speed data.

$$q = \frac{u}{c_1 + \frac{c_2}{u_f - u} + c_3 u} \tag{15}$$

$$c_1 = \frac{u_f}{k_j u_c^2}(2u_c - u_f); \quad c_2 = \frac{u_f}{k_j u_c^2}(u_f - u_c)^2; \quad c_3 = \frac{1}{q_c} - \frac{u_f}{k_j u_c^2} \tag{16}$$

**CASE STUDY**

Two data sets are used in the case study to validate the performance of the proposed method. The first data set is the loop detector data along Interstate 5 (I-5) in Portland, which is used to validate the accuracy of binary result from congestion identification in the proposed method. The second data set is the INRIX data along Interstate 66 (I-66) in Northern Virginia, which is used to test the proposed method of extracting bottleneck characteristics and bottleneck visualization.

The loop detector data consisted of 24 days' worth of high quality data: midweek, non-holiday days between February and December of 2008. The data were collected from archived data from the northbound I-5 corridor in Portland, Oregon, metropolitan region. The freeway corridor is 22 miles (35 kilometers) long, and includes 22 loop detectors as illustrated in FIGURE 6 (a). Considering the poor data quality for two detectors, we only used the date collected by 20 detectors. The raw data collected by every 20 seconds between 5:00 am and 10:00 pm were aggregated by five-minute interval to remove measurement error and data noise. So the daily speed data is a matrix with dimension 20 by 204.

The INRIX data collected during the year of 2013 along I-66 eastbound are also included in the test data sets, giving that major traffic bottlenecks are usually observed in this corridor. The selected freeway corridor on I-66 is presented in FIGURE 6 (b), which includes 36 freeway sections (the start location of each section is highlighted) with the total length of 30.7 miles. Segment-based average speeds are provided in the raw data, which are collected by every one minute. The raw speed data are also aggregated by five-minute interval. Thus, the traffic speed matrix over spatial (upstream to downstream) and temporal (from 0:00 am to 23:55 pm) domains can be obtained for each day, which is a data matrix with dimension 36 by 288.

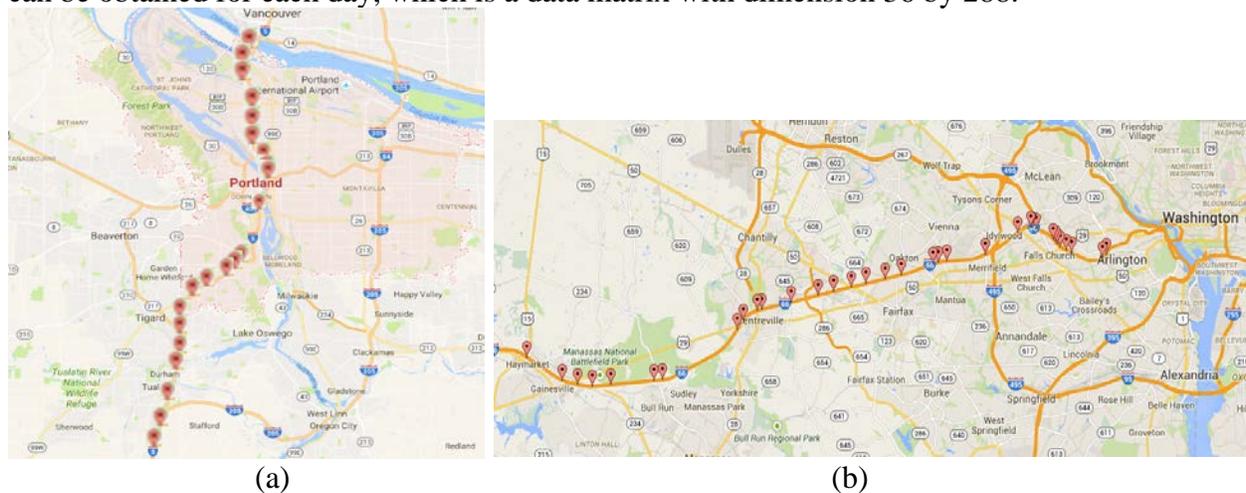

(a)          (b)
**FIGURE 6. Layout of the selected freeway corridors. (a) I-5. (b) I-66.**

**Evaluation of Binary Result by Congestion Identification**
The performance of using congestion identification to obtain binary result is tested with the Portland loop detector data because it contains the ground truth data, which is defined by a manual procedure. In this manual procedure, the activation and deactivation times for each candidate bottleneck are carefully diagnosed and verified with oblique curves of cumulative vehicle arrival versus time and cumulative occupancy (or speed) versus time, constructed from data measured at neighboring freeway loop detectors [22-24]. The ground truth data obtained for each day is a binary matrix, where 1 represents a congestion roadway section and 0 represents a free flow section. Though the available ground truth data obtained by this procedure is not accurate because it depends on the estimated free flow speed, it's the only available reference so we use it as the ground truth data to validate the binary results by congestion identification.

To better evaluate the performance of congestion identification in the proposed method, the Chen's algorithm [6] is also tested in the same data set to generate binary results and compare with the proposed method. The Chen's algorithm automatically identifies speed data into either congested or uncongested traffic condition using loop detector data. Specifically, this method compares speed data from each pair of neighboring detectors and determines the existence of bottleneck (recognized as congested traffic condition) when
- Speed at the upstream detector is below the maximum speed threshold $u_{max}$.
- Difference in the speeds at the upstream and downstream detectors is above the minimum speed differential threshold $\Delta u_{min}$.

Different combinations of parameter sets are evaluated in [25] for the Portland data set, and the optimum parameters for five-minute data aggregation level are selected as $u_{max} = 35$ mph, $\Delta u_{min} = 15$ mph. In this test, the proposed congestion identification method consists of the procedures of image binarization and post-process filters. The parameters are set as $\alpha_1 = 2$, $\alpha_2 = 3$, and $\alpha_3 = 20$, which means that a structure of 2 roadway sections by 3 time intervals (15 minutes) is used in the morphology opening and closing operations to filter noises and the connected areas with less than 20 cells are deleted as localized congested regions.

For each data point in the daily speed matrix, the automated methods including the proposed method and the Chen's algorithm can determine either it's congested or uncongested traffic condition. Meanwhile, the ground truth data tells whether this data point was truly congested or not. Therefore, all the test data points can be categorized as true positives (correctly identified "congested"), false positives (identified as "congested" but are actually "uncongested"), true negatives (correctly identified "uncongested"), and false negatives (identified as "uncongested" but are actually "congested"). Then the success rate and the false alarm rate can be computed.

The success rate and the false alarm rate of congestion identification are calculated for each day by using the Chen's algorithm and the proposed method. FIGURE 7 illustrates the success and false alarm rates by the Chen's algorithm sorted in an ascending order and the corresponding results by the proposed method. The plots indicate that the proposed method clearly outperforms the Chen's algorithm to produce higher success rate and lower false alarm rate, which means more congested speed data are correctly recognized as "congested" and less uncongested data are incorrectly recognized as "congested" in the proposed method.

$$\text{Success Rate} = \frac{\text{True Positive}}{\text{True Positive} + \text{False Negative}} \quad (17)$$

$$\text{False Alarm Rate} = \frac{\text{False Positive}}{\text{True Positive} + \text{False Positive}} \quad (18)$$

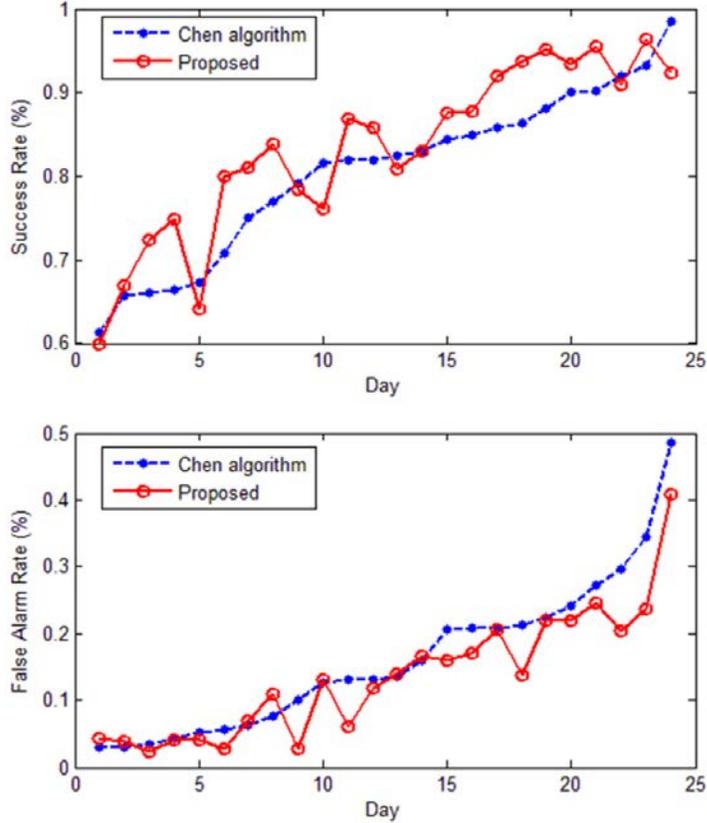

**FIGURE 7. Accuracy of binary results from congestion identification using Chen's algorithm versus the proposed method.**

**Evaluation of Bottleneck Identification and Visualization**
The INRIX data along I-66 in the Northern Virginia region are used to evaluate the proposed bottleneck identification method to extract bottleneck characteristics (activation/deactivation points, shockwave speed and estimated delay caused by bottleneck) and bottleneck visualization. Note that we cannot quantify the accuracy of the extracted bottleneck characteristics since the ground truth data are not available. Thus, the proposed method is implemented to the daily speed matrix in the INRIX data set and the results are visually accessed. The parameters are set as $\alpha_1 = 2$, $\alpha_2 = 3$, $\alpha_3 = 20$, $\lambda_1 = 1.3$, $u_f = 60$ mph, $u_c = 45$ mph, $k_j = 150$ veh/km/lane, $q_c = 2500$ veh/hour. The maximum shockwave speed is calculated as 14.6 mph by following the constraint in (12).

The testing results demonstrate the proposed method can effectively and accurately extract and visualize the major characteristics of freeway bottlenecks. FIGURE 8 demonstrates the samples of bottleneck identification and visualization using the proposed method for three days. Each isolated bottleneck is automatically separated and represented by a different color in FIGURE 8 (b). The activation and deactivation points on the front and rear of bottlenecks are extracted in FIGURE 8 (c). The slope of white line connecting two activation points indicates the speed for back-propagation shockwave. Table I presents the extracted bottleneck characteristics for speed matrix on June 17 2013. Totally seven bottlenecks are identified. For each isolated bottleneck, the activation/deactivation points on the front and rear of bottleneck are denoted by the section id and time stamp. Thereafter, the back-propagation shockwave speed can be calculated using the front and rear activation points. Considering the maximum shockwave speed is 14.6 mph, the calculated shockwave speeds of 15.8 mph and 28.9 mph for the second and fourth bottlenecks are identified as incorrect values. So we adjusted both values as 14.6 mph. Besides, the total delay caused by each bottleneck can be estimated using (13) ~ (15). We can see that the third and sixth bottlenecks cause tremendous impacts on freeway traffic with the estimated delay of 1518 and 1398 vehicle hours respectively. Note that sometimes the bottleneck identification results by the proposed method are not very accurate. For instance, two bottlenecks are not correctly separated for the speed matrix on June 19 2013.

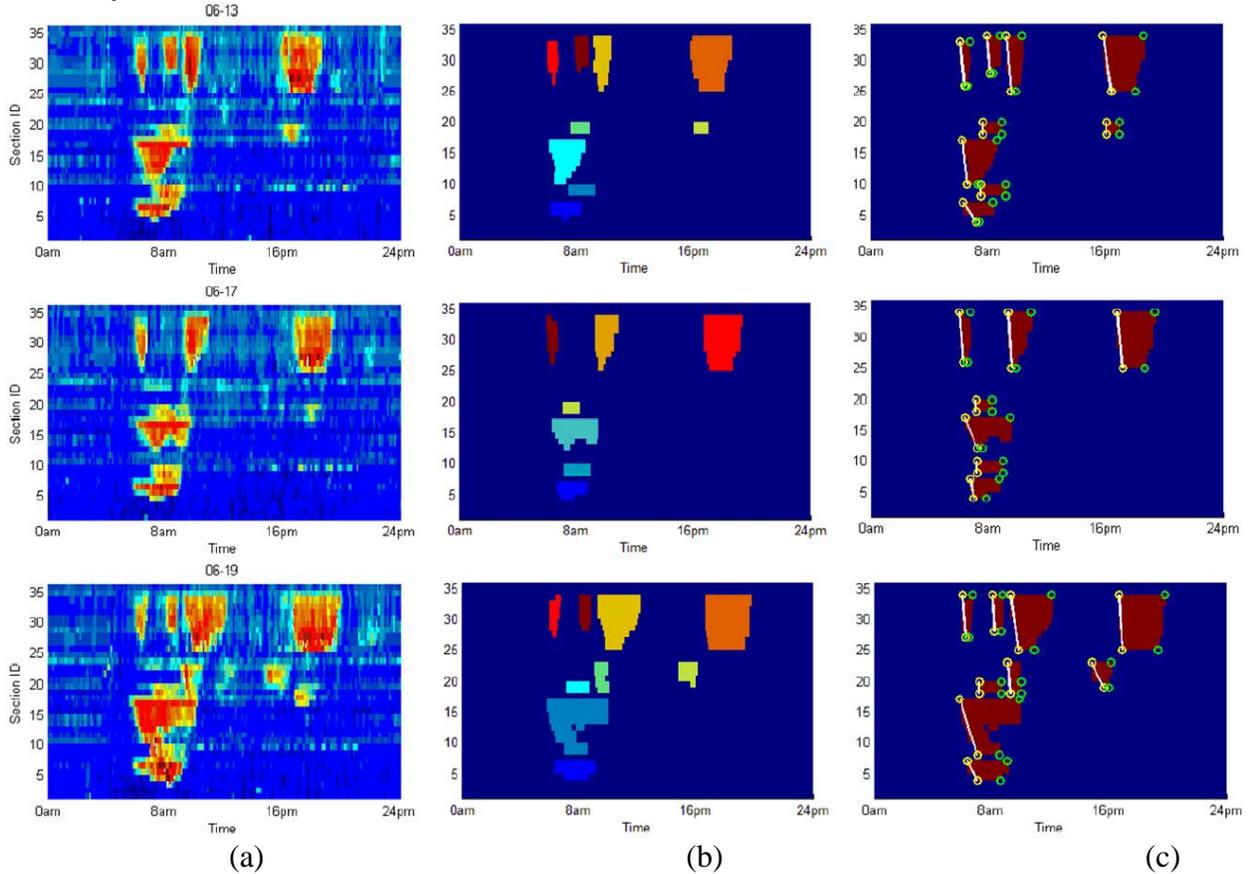

**FIGURE 8. Samples of bottleneck identification and visualization. (a) speed contour. (b) isolated bottlenecks by different colors. (c) activation and deactivation points, and shockwave speed.**

**TABLE 1. Extracted bottleneck characteristics for speed matrix on June 17 2013**

| Bottleneck Index | Front activation point [section id, time] | Front deactivation point [section id, time] | Rear activation point [section id, time] | Rear deactivation point [section id, time] | Shockwave speed (mile per hour) | Estimated delay (vehicle hour) |
|---|---|---|---|---|---|---|
| 1 | 6, 6:45 am | 6, 8:50 am | 4, 7:05 am | 4, 7:50 am | 10.2 | 489 |
| 2 | 9, 7:15 am | 9, 9:05 am | 8, 7:20 am | 8, 9:00 am | **15.8** (14.6) | 344 |
| 3 | 16, 6:25 am | 16, 9:35 am | 12, 7:20 am | 12, 7:35 am | 4.1 | 1518 |
| 4 | 19, 7:10 am | 19, 8:20 am | 18, 7:15 am | 18, 8:15 am | **28.9** (14.6) | 274 |
| 5 | 33, 9:20 am | 33, 11:00 am | 25, 9:40 am | 25, 9:55 am | 9.1 | 854 |
| 6 | 33, 16:45 pm | 33, 19:25 pm | 25, 17:15 pm | 25, 18:40 pm | 6.2 | 1398 |
| 7 | 33, 6:00 am | 33, 6:50 am | 26, 6:25 am | 26, 6:35 am | 5.5 | 235 |

## CONCLUSIONS

This paper develops an automatic freeway bottleneck identification algorithm by combining image processing techniques and traffic flow theory. In the proposed framework, the raw spatiotemporal speed data are transformed into binary matrices using image binarization techniques. Then two post filters are proposed to further clean up the binary matrices by filtering salt-and-pepper noises and localized congested regions. Subsequently, roadway geometry information is used to remove the impact of the acceleration area downstream of a bottleneck. Finally, the major characteristics of the bottleneck including activation and deactivation points, shockwave speeds and traffic delay caused by the bottleneck are extracted and visualized. Two case studies using loop detector data on I-5 and INRIX probe data on I-66 are conducted. The test results demonstrate that the proposed method can accurately and effectively identify bottlenecks. In future work, more field data from different freeway corridors and various sensing technologies will be used to validate the performance of the proposed algorithm. In addition, further development of the model to enhance identification accuracy will be needed for the future research.

## ACKNOWLEDGMENT
The authors thank Professor Robert Bertini for providing the Portland loop detector data and the ground truth binary results that were used in the case study. This work was funded by the Department of Energy through the Office of Energy Efficiency and Renewable Energy (EERE), Vehicle Technologies Office, Energy Efficient Mobility Systems Program under award number DE-EE0008209.